\documentclass[11pt]{article}
\usepackage{geometry}             
\usepackage{graphicx}
\usepackage{amssymb}
\usepackage{amsmath}
\usepackage{epstopdf}
\usepackage{cite}

\usepackage[hyperindex=true,
          pdfstartview=FitH,
          bookmarksnumbered=true,
          bookmarksopen=true,
          citecolor=blue,
          linkcolor=blue,
          colorlinks=true,
          unicode]{hyperref}

\usepackage[cal=boondox,
bb=ams,
scr=esstix
]{mathalfa}
\newcommand{\ms}[1]{\mathscr{#1}}
\newcommand{\df}[2]{\frac{\rd{#1}}{\rd{#2}}}

\allowdisplaybreaks

\allowdisplaybreaks

\newcommand{\rd}{{\rm d}}
\newcommand{\re}{{\rm e}}

\newcommand{\pfrac}[2]{\left(\frac{\partial #1}{\partial #2}\right)}
\newcommand{\bfrac}[2]{\left(\frac{#1}{#2}\right)}

\newcommand{\blue}[1]{\textcolor{blue}{#1}}
\renewcommand{\blue}[1]{\textcolor{black}{#1}}  

\begin{document}

\title{High temperature AdS black holes are \\
low temperature quantum phonon gases}
\author{Xiangqing Kong\thanks{
{\em email}: \href{mailto:2120200165@mail.nankai.edu.cn}{2120200165@mail.nankai.edu.cn}},~
Tao Wang\thanks{
{\em email}: \href{mailto:taowang@mail.nankai.edu.cn}{taowang@mail.nankai.edu.cn}},~
and Liu Zhao\thanks{Correspondence author, {\em email}: 
\href{mailto:lzhao@nankai.edu.cn}{lzhao@nankai.edu.cn}}\\
School of Physics, Nankai University, Tianjin 300071, China
}

\date{}
\maketitle

\begin{abstract}
We report a precise match between the high temperature $(D+2)$-dimensional 
Tangherlini-AdS black hole and the low temperature quantum phonon gas
in $D$-dimensional nonmetallic crystals residing in $(D+1)$-dimensional flat spacetime. 
The match is realized by use of the recently proposed restricted phase space 
formalism for black hole thermodynamics, and the result can be viewed as a 
novel contribution to the AdS/CMT  correspondence on a quantitative level.

\vspace{1em}


\end{abstract}

\section{Introduction}

Black holes are the most important objects predicted by the modern relativistic 
theories of gravity. Since the 1970s it has been commonly believed that black holes 
are thermal objects\cite{bekenstein1972black, bekenstein1973black, bardeen1973four, 
hawking1975particle, bekenstein1975statistical}, and as such they must contain a 
large number of microscopic degrees
of freedom, and understanding the nature of these microscopic degrees 
of freedom might provide a window for inspecting the quantum nature 
of gravitation. Although the goal for 
understanding quantum gravity is still far beyond the scope of our human's sight, 
there indeed has been a number of 
progresses and speculations toward the understanding of the microscopic structure of 
black holes, among which the most notable ideas come from the AdS/CFT 
correspondence\cite{hooft1993dimensional,susskind1995world,maldacena1998large} .

In the recent years, another line of thinking pumps up which attempts to 
understand the black hole microstructure purely from thermodynamic perspective, 
especially following the so-called extended phase space (EPS) formalism 
\cite{kastor2009enthalpy, 
dolan2010cosmological, dolan2011pressure, dolan2011compressibility, kubizvnak2012p, 
cai2013pv, kubizvnak2017black, xu2014critical, xu2014gauss, lemos2018black, 
zhang2015phase}. 
The EPS formalism to black hole thermodynamics is an approach which takes the 
negative cosmological constant as a thermodynamic variable which is proportional 
to the pressure and is accompanied by an associated conjugate variable known as 
the thermodynamic volume. A significant amount of works have been published 
about the behavior of various black holes under the EPS approach, mostly concentrated in the
critical phase transitions. Some authors also attempted to explore the interaction potential
between the microscopic degrees of freedom (also called black hole molecules) 
of the black holes starting from the 
thermal equation of states that arises from the EPS 
formalism \cite{wei2015insight,wei2020extended,dehyadegari2020microstructure}.

Since the end of the last year, we proposed and kept working on an alternative formalism 
of black hole thermodynamics called the restricted phase space (RPS)
formalism \cite{gao2021restricted, 
gao2022thermodynamics, wang2021black, zhao2022thermodynamics}. 
The RPS formalism differs from the EPS formalism in that the cosmological constant is 
no longer taken as a thermodynamic variable but rather kept as a constant. Instead, we 
allow the gravitational constant $G$ to be variable and introduced a pair of new 
thermodynamic quantities $N= L^D/G, \mu= GT I_E/L^D$ which are respectively interpreted 
as the effective number of microscopic degrees of freedom
(or alternatively the number of black hole molecules) and the chemical potential 
of the black hole, wherein $D$ is the dimension of the black hole bifurcation horizon, 
\blue{$I_E$ is the Euclidean action of the black hole spacetime and $T$ is the black 
hole temperature. The constant length scale $L$ corresponds to the 
maximum value for the radius of the event horizon in any thermodynamic 
processes of interests, which can be chosen arbitrarily besides the above requirement. 
The arbitrariness of $L$ may be fixed in several ways, e.g. by taking some specific 
convention for the unit of the chemical potential, or by comparison to low 
temperature quantum phonon gases when considering the high temperature limit. 
We shall come back on this last possibility later.}
  
The RPS formalism is a restricted version of the holographic thermodynamics for AdS/CFT 
systems proposed by Visser\cite{visser2022holographic}, 
but with the removal of cosmological constant from the 
list of thermodynamic variables. This removal of cosmological constant resolves a 
number of issues, including, but not limited to, the following points: 1) The Euler 
homogeneity holds perfectly without introducing rational coefficients in the mass formula;
2) The theory-changing problem which we call the ensemble of theories issue which existed in
the EPS formalism is completely avoided, and the black hole mass restores {its} original 
interpretation as internal energy rather than as enthalpy as did in the EPS formalism; 3) 
The definition of $N, \mu$ are now independent of the holographic duality, and thus the 
RPS formalism applies perfectly to the cases of non-AdS black holes. \blue{There are some 
arguments \cite{Gregory} which can avoid the theory-changing problem in the presence of 
variable cosmological constant, but the lack of complete Euler homogeneity and the 
inability for generalizations to non-AdS cases remain to be critical issues in other 
existing formalisms.}

Subsequent works \cite{kong2022restricted, Sadeghi, bwu} revealed that the RPS 
formalism works perfectly for black holes with different asymptotics in 
Einstein gravity and certain higher curvature 
gravity models in diverse spacetime dimensions. 
Some universal behaviors have also been found for black holes under this formalism, 
e.g. the black holes in Einstein-Hilbert  and Born-Infeld like gravities behave 
qualitatively the same, while black holes in Chern-Simons like gravities behave
completely different. Moreover, for certain AdS black holes, the high temperature 
limit of the heat capacity has a power law dependence on the temperature which is 
identical to the low temperature limit of the Debye heat capacity of nonmetallic 
crystals. We believe that this similarity is not a coincidence, and this letter is
a further exploration on this similarity. As will be shown in the main text below, 
the similarity between the high temperature AdS black holes and the low temperature 
nonmetallic crystals is much more profound. Not only the heat capacities, but also 
the Helmholtz free energies, the internal energies and the entropies of the two drastically 
different types of systems (high temperature AdS black hole and low temperature 
nonmetallic crystals) behave almost identically in certain temperature ranges. 
Since the microscopic description of 
nonmetallic crystal is identical to a quantum phonon gas, we conclude that the 
high temperature AdS black holes are actually equivalent to low temperature 
quantum phonon gases. Although this link still does not reveal the nature of individual 
black hole molecules, it indeed reveals their collective effects, i.e. the black hole 
molecules are quantum, and their collective motion behaves like phonons 
in certain temperature limit. We believe 
that this observation is a meaningful progress towards the ultimate understanding 
of the black hole microstructure as well as the quantum nature of gravitation.

\section{Thermodynamics of Tangherlini-AdS black holes and the high temperature limit}

We will exemplify the connection between high temperature AdS black holes and 
low temperature Bose gases by studying in detail the thermodynamics of 
Tangherlini-AdS black holes in the RPS formalism. 
Throughout this letter we work in units $c=1, k_B=1, \hbar=1$ but keep the Newton constant 
$G$ intact.

We take the spacetime dimension to be $D+2$, so that the bifurcation horizon for 
the black hole is $D$. The $(D+2)$-dimensional Einstein-Hilbert action
with Gibbons-Hawking boundary term is given by
\begin{align}
I=\frac{1}{16\pi G}\int_{\mathcal M} (R-2\Lambda)\sqrt{g}\,\rd^{D+2}x 
+\frac{1}{8\pi G}\int_{\partial\mathcal{M}} K \sqrt{h}\,\rd^{D+1}x,
\end{align}
where $g=|{\rm det}(g_{\mu\nu})|$, $h=|{\rm det}(h_{ab})|$, with $h_{ab}$ being 
the induced metric on the boundary $\partial\mathcal{M}$ of the spacetime $\mathcal M$
and $K$ being the trace of the extrinsic curvature of $\partial\mathcal{M}$ 
in $\mathcal M$. The inclusion of the Gibbons-Hawking boundary term is important in order to
obtain the correct value for the Euclidean action $I_E$.

The metric of the $(D+2)$-dimensional Tangherlini-AdS black hole written in 
spherical coordinates takes the form
\begin{align}
\rd s^2&= -f(r)\rd t^2+f^{-1}(r)\rd r^2+r^2\rd\Omega_D^2,\\
f(r)&=1-\frac{16\pi G}{D\mathcal{A}_D}\frac{M}{r^{D-1}}+\frac{r^2}{\ell^2},
\end{align}
where $\rd\Omega_D^2$ is the line element on a unit $D$-sphere with area  
$\mathcal{A}_D=\frac{2\pi^{(D+1)/2}}{\Gamma\left(\frac{D+1}{2}\right)}$, and 
$\ell$ is the AdS radius which is related to the negative cosmological constant via
\[
\Lambda=-\frac{D(D+1)}{2\ell^2}.
\] 

The black hole event horizon is located at $r=r_h$ which is a root of the equation 
$f(r)=0$. Accordingly, the mass of the black hole can be expressed 
in terms of $r_h$ and $G$ as
\begin{align}
M=\frac{D\mathcal{A}_D r_h^{D-1}}{16\pi G}\left(1+\frac{r_h^2}{\ell^2}\right)
=\frac{D \pi ^{(D-1)/2} r_h^{D-1} \left(\ell^2+r_h^2\right)}
{8 \ell^2 \Gamma \left(\frac{D+1}{2}\right) G}.
\end{align}

In the RPS formalism, the Tangherlini-AdS black hole has two independent 
extensive variables, i.e. the entropy $S$ and the effective number of microscopic 
degrees of freedom $N$, 
\begin{align}
S&=\frac{\mathcal{A}_{D}r_h^{D}}{4G} 
= \frac{\pi ^{(D+1)/2}  r_h^{D} }{2 \Gamma \left(\frac{D+1}{2}\right)G},\\
N&=\frac{L^D}{G}, \label{NG}
\end{align}
where $r_h$ is the radius of the event horizon of the black hole which is a 
root of the equation $f(r)=0$. The corresponding intensive variables are
\begin{align}
T&=\frac{1}{4\pi} \pfrac{f}{r}_{r=r_h}
= \frac{(D-1) \ell^2+(D+1) r_h^2}{4 \pi  \ell^2 r_h},
\label{Tr}\\
\mu&= \frac{GT I_E}{L^D}= \frac{\pi ^{(D-1)/2} r_h^{D-1} \left(\ell^2-r_h^2\right)}
{8 \ell^2  L^D  \Gamma \left(\frac{D+1}{2}\right)},
\end{align}
\blue{where the Euclidean on-shell action $I_E$  can be calculated following the 
standard procedures as given in, e.g. \cite{G, York, Chamblin:1999tk, Gibbons2}.
} 

As expected, the two intensive variables are not independent of each other, because
the first law
\[
\rd M=T\rd S+\mu\rd N
\]
and the Euler homogeneity relation
\begin{align}
M=TS+\mu N. \label{euler}
\end{align}
hold simultaneously, which implies the Gibbs-Duhem relation
\[
S\rd T+N\rd \mu=0.
\]
\blue{The complete Euler homogeneity is an outstanding feature of the RPS formalism
in contrast to alternative proposals, see, e.g. \cite{visser2022holographic,Cong1},  
which included additional variables $P,V$, with the sacrifice of complete Euler
homogeneity ($M$ is first order homogeneous in $S, N$ but not in $V$). 
The framework introduced in \cite{visser2022holographic} could be meaningful 
for the dual CFT in the context of holographic duality, however, in the absence of 
complete Euler homogeneity, it may cause some problems in understanding black 
hole phase transitions in a scale independent way, because the law of corresponding 
states cannot be realized in the presence of $P-V$ variables. In the RPS formalism, 
the law of corresponding states holds perfectly, making the $T-S$ phase transitions 
scale independent, as we have shown in Refs.\cite{gao2021restricted, gao2022thermodynamics, 
kong2022restricted} for different black hole solutions. Please be reminded that 
scale independence is a characteristic property that any phase transition should 
have. }

\blue{Let us proceed to analyze the high temperature limit for the 
Tangherlini-AdS black hole.} 
It is preferable to replace the geometric 
parameter $r_h$ and the coupling {coefficient} $G$ in the above expressions for 
thermodynamic quantities by a set of independent macro state parameters. 
The standard practice is to take the extensive parameters $(S,N)$ 
as independent variables and re-express
$M$ (understood as the internal energy) and the intensive variables $T,\mu$ as functions 
in $(S,N)$: $M=M(S,N), T=T(S,N), \mu=\mu(S,N)$. 
However, since the major goal of this work is to analyze the high temperature limit
of various thermodynamic quantities, we prefer to take $(T,N)$ as independent variables
and rewrite the other macro state functions in terms of these. From eqs.\eqref{NG} and
\eqref{Tr} it is straightforward to get 
\begin{align}
G&=\frac{L^{D}}{N},\\
r_h&=\frac{\ell \left(2 \pi  \ell T \pm \sqrt{4 \pi ^2 \ell^2 T^2-D^2+1}\right)}{D+1}.
\label{rh}
\end{align}
The two branched values for $r_h$ indicate that there are two black hole states
at the same temperature, of which the smaller one (i.e. the negative branch) is 
unstable because it has smaller entropy. One can easily check that, in the 
high temperature limit, the value of the negative branch of $r_h$ goes to zero, 
\blue{and so are the mass and entropy}. Therefore, in order to study the high 
temperature limit, we only need to consider the positive branch of $r_h$. 
In this branch, the entropy and the internal energy of 
the black hole can be rewritten as
\begin{align}
S&=\frac{N \pi ^{(D+1)/2}}{2 \Gamma \left(\frac{D+1}{2}\right)}
\left(\frac{\ell \left(\sqrt{4 \pi ^2 \ell^2 T^2-D^2+1}+2 \pi \ell T\right)}
{(D+1) L}\right)^{D},\\
M&=\frac{N D \pi ^{(D-1)/2} \left(\frac{\ell 
\left(\sqrt{4 \pi ^2 \ell^2 T^2 -D^2 +1}+2 \pi  \ell T\right)}{D+1}\right)^{D-1} 
\left(\frac{\left(\sqrt{4 \pi ^2 \ell^2 T^2-D^2+1}+2 \pi  \ell T\right)^2}
{(D+1)^2}+1\right)}{8 L^{D} \Gamma \left(\frac{D+1}{2}\right)},
\end{align}
\blue{both tend to be divergent at $T\to\infty$. Notice that this branch of states
is specific to AdS black holes and does not exist for asymptotically flat and de Sitter
cases.} The Helmholtz free energy $F$ and the heat capacity $C_N$ of the black hole
can be written as
\begin{align}
F&= M-TS\nonumber\\
&= \frac{N \pi ^{(D-1)/2}  
\left(\frac{\ell \left(\sqrt{4 \pi ^2 \ell^2 T^2 -D^2 +1}+2 \pi  \ell T\right)}
{D+1}\right)^{D-1} \left(1-\frac{\left(\sqrt{4 \pi ^2 \ell^2 T^2 -D^2 +1}
+2 \pi  \ell T\right)^2}{(D+1)^2}\right)}{8 L^{D}\Gamma \left(\frac{D+1}{2}\right)},\\
C_N&=T\pfrac{S}{T}_N=
\frac{N D \pi ^{(D+3)/2} \ell T \left(\frac{
\ell \left(\sqrt{4\pi ^2 \ell^2 T^2-D^2+1}+2 \pi  \ell T\right)}
{(D+1) L}\right)^{D}}{\Gamma
\left(\frac{D+1}{2}\right) \sqrt{4 \pi ^2 \ell^2 T^2-D^2+1}}.
\end{align}

The temperature dependence of the above results appears to be very complicated. 
However, if we consider the high temperature limit $T\to \infty$, the results
become much simplified,
\begin{align}
\lim_{T\to\infty}M &= \frac{N  \pi ^{1/2}D}{2(D+1)\Gamma \left(\frac{D+1}{2}\right)} 
\bfrac{\pi^{3/2}(2\ell)^2}{(D+1)L}^DT^{D+1},\\
\lim_{T\to\infty}F &= -\frac{N  \pi ^{1/2}}{2(D+1)\Gamma \left(\frac{D+1}{2}\right)} 
\bfrac{\pi^{3/2}(2\ell)^2}{(D+1)L}^DT^{D+1},\\
\lim_{T\to\infty}S &= \frac{N  \pi ^{1/2}}{2\Gamma \left(\frac{D+1}{2}\right)} 
\bfrac{\pi^{3/2}(2\ell)^2}{(D+1)L}^D T^D,\\
\lim_{T\to\infty}C_N &=\frac{N  \pi ^{1/2}D}{2\Gamma \left(\frac{D+1}{2}\right)} 
\bfrac{\pi^{3/2}(2\ell)^2}{(D+1)L}^D T^D.
\end{align} 

At this point we need to make it clear what is meant by the high temperature limit 
$T\to\infty$. Since $T$ is a dimensionful quantity, it does not make sense to say the
temperature is high or low without comparing to a constant characteristic 
temperature. It is evident that in the present case, the characteristic 
temperature $T_{\rm{bh}}$ can be chosen as
\[
T_{\rm{bh}}= \frac{(D+1)L}{\pi^{3/2}(2\ell)^2}.
\]
With this choice we can rearrange the above high temperature (i.e. $T\gg {T_{\rm{bh}}}$) 
values of thermodynamic quantities in the form
\begin{align}
M&\approx \frac{\pi ^{1/2}}{2(D+1)\Gamma \left(\frac{D+1}{2}\right)} 
D N T \bfrac{T}{T_{\rm{bh}}}^D,
\label{Mbh}\\
F&\approx -\frac{\pi ^{1/2}}{2(D+1)\Gamma \left(\frac{D+1}{2}\right)} 
N T \bfrac{T}{T_{\rm{bh}}}^D,\\
S&\approx \frac{\pi ^{1/2}}{2\Gamma \left(\frac{D+1}{2}\right)} 
N\bfrac{T}{T_{\rm{bh}}}^D,\\
C_N&\approx \frac{\pi ^{1/2}}{2\Gamma \left(\frac{D+1}{2}\right)} 
D N\bfrac{T}{T_{\rm{bh}}}^D.\label{CNbh}
\end{align}

\section{Relation to low temperature phonon gases}

The power law temperature dependence of the black hole thermodynamic 
quantities $M,F,S,C_N$ reminds us of the quantum phonon gases that appear in 
nonmetallic crystals. Let us recall the following known results \cite{zhaotherm}
for the internal energy $E$, Helmholtz free energy $F$, entropy $S$ and 
isochoric heat capacity $C_V$ of the $D$-dimensional phonon gases 
in nonmetallic crystals residing in $(D+1)$-dimensional flat 
spacetime, which can be obtained straightforwardly by use of (grand)canonical 
ensemble and the Debye's linear dispersion relation $\epsilon(k)=v_s|k|$ for phonons,
\begin{align}
E&=DN T\ms{D}_D\bfrac{T_{\mathrm{D}}}{T},\\
F&=DN T\log\left(1-\re^{-T_{\mathrm{D}}/T}\right) 
- NT \ms{D}_D\bfrac{T_{\mathrm{D}}}{T},\\
S&=-DN \log\left(1-\re^{-T_{\mathrm{D}}/T}\right) +(D+1)N
\ms{D}_D\bfrac{T_{\mathrm{D}}}{T},\\
C_V&=DN \ms{L}_D\bfrac{T_{\mathrm{D}}}{T},
\end{align}
where $N$ is the number of crystal lattice atoms, $T_D$ is the Debye temperature, 
$\ms{D}_D(x)$ is 
the $D$-dimensional Debye function
\[
\ms{D}_D(x)\equiv D x^{-D} \int_0^x \frac{y^D}{\re^y-1}\rd{y},
\]
and $\ms{L}_D(x)$ is the $D$-dimensional Langevin function,
\[
\ms{L}_D(x)\equiv \ms{D}_D(x)-x \df{}{x}\ms{D}_D(x).
\]
At low temperature $T\ll T_D$, the above phonon gas functions behave as
\begin{align}
E &\approx f(D) D N T\bfrac{T}{T_{\mathrm{D}}}^D, 
\label{Eph}\\
F &\approx -f(D) N T\bfrac{T}{T_{\mathrm{D}}}^D,\\
S &\approx f(D)(D+1)N\bfrac{T}{T_{\mathrm{D}}}^D, \\
C_V &\approx  f(D) D(D+1)N\bfrac{T}{T_{\mathrm{D}}}^D,
\label{CVph}
\end{align}
where $f(D)=D\zeta(D+1)\Gamma(D+1)$.

We can see a surprising similarity between the high temperature 
results \eqref{Mbh}-\eqref{CNbh} for black holes and the low temperature results
\eqref{Eph}-\eqref{CVph} for quantum phonon gases.
In fact, if we make the identification
\begin{align}
T_{\rm D}= \bfrac{2(D+1)\Gamma \left(\frac{D+1}{2}\right)f(D)}{\pi^{1/2}}^{1/D} 
T_{\rm bh},
\label{tteq}
\end{align}
then there will be a precise quantitative match between eqs. 
\eqref{Mbh}-\eqref{CNbh} and eqs. \eqref{Eph}-\eqref{CVph}, 
provided we further identify $M$ with $E$ and $C_N$ with $C_V$.
We can even absorb the constant factor appearing on the right hand side of 
eq.\eqref{tteq} by a simple redefinition of the length scale $L$,
\[
L\to \bfrac{2(D+1)\Gamma \left(\frac{D+1}{2}\right)f(D)}{\pi^{1/2}}^{1/D} L,
\]
which is allowed because of the arbitrariness in the choice of $L$. 
Then the Debye temperature $T_{\mathrm{D}}$ for the phonon gas will be identical to the 
black hole characteristic temperature $T_{\rm bh}$. 
We thus conclude that the Tangherlini-AdS black holes at high temperature are
nothing but quantum phonon gases in nonmetallic crystals at low temperature. 
In this correspondence, the number of black hole molecules is identified with the
number of crystal lattice atoms which are both denoted as $N$.

\section{Concluding remarks and discussions}

The major conclusion of this work can be summarized in a single sentence: 
{\it Tangherlini-AdS black holes at high temperature are equivalent to 
quantum phonon gases in nonmetallic crystals at low temperature.}
Since we have already {evidence} about the same power law dependences of heat capacities 
for different black hole solutions in different gravity models\cite{kong2022restricted}, 
it is natural to expect that this {\em AdS/phonon gas correspondence} may also hold for 
other asymptotically AdS black holes. The AdS asymptotics is a necessary 
condition for the above correspondence to hold, because for non-AdS black holes 
the heat capacities are identically negative \cite{zhao2022thermodynamics}, 
which could not have direct relationship to normal phonon gases.

Asymptotically AdS black holes have been extensively studied during the past 25 years
or so, mostly in connection with AdS/CFT correspondence and its various extensions
such as AdS/CMT (condensed matter theory) 
\cite{hartnoll2009lectures,sachdev2011condensed,zaanen2015holographic}, 
AdS/QCD \cite{karch2006linear,kajantie2007thermodynamics,panero2009thermodynamics}, 
etc. Although a countless number of theoretical results have been obtained 
in these fields, most of the correspondences remain 
qualitative. The present work adds some extra contribution to the field of AdS/CMT
correspondence with a precise quantitative match between the two sides.
Let us remark that the AdS/phonon gas correspondence described in this work 
holds in generic spacetime dimension $D+2$, where $D$ is the dimension
of the bifurcation horizon of the black hole and $D+1$ is the dimension of the 
boundary of the black hole spacetime in which the $D$-dimensional 
nonmetallic crystal resides. Let us stress that the AdS/phonon gas correspondence 
that we report here is established purely on the level of thermodynamics. No attempts 
on the microscopic description were made here, and there is no need to assume 
broken translational symmetries from the beginning, as did in previous works 
on similar topics in the context of AdS/CFT correspondence \cite{ling2014metal,
cai2017intertwined, alberte2018black,alberte2018holographic}. 
On the level of thermodynamics, the correspondence is complete, because, according to the 
famous Massieu's theorem, the thermodynamic properties of 
any macroscopic system are completely determined by a single thermodynamic potential 
with adapted independent variables. In the present context, the Helmholtz free 
energy $F=F(T,N)$ plays the role of the thermodynamic potential on both sides.

From the point of view of statistical physics, the power law dependence of thermodynamic
quantities for the phonon gas is solely determined by Debye's linear dispersion relation. 
The AdS/phonon gas correspondence seems to indicate that the dispersion relation 
for black hole molecules at high temperature is also linear, 
provided the underlying statistical physics principles are the same. Although the 
black hole microstructure remains unclear as we still cannot say anything about 
individual black hole molecules,  the result of the present work does 
reveal that the black hole molecules are quantum, and collectively behave as 
phonons at sufficiently high temperature. 

One may wonder why the AdS/phonon gas correspondence could hold, 
given that the AdS black hole 
has a chemical potential whilst the phonon gas has not. This is actually a 
misreading of the role of black hole chemical potential. 
As we have already pointed out 
in the end of the last section, the number of black hole molecules corresponds to
the number of crystal lattice sites, rather than the number of phonons --- the latter
is not conserved, and hence the phonon gas has a vanishing chemical potential. On the
other hand, the number of lattice atoms is conserved and hence the crystal background
does have a nonvanishing chemical potential which is connected with the binding energy 
of the crystal and is responsible for the lattice growth. Likewise,
the black hole chemical potential should also be connected with the binding 
energy of the black hole. Using the Euler homogeneity relation \eqref{euler},
it is straightforward to write down the high temperature behavior of the 
black hole chemical potential, 
\begin{align}
\mu&= \frac{F}{N} \approx -\frac{\pi ^{1/2}}
{2 (D+1) \Gamma \left(\frac{D+1}{2}\right)} T\bfrac{T}{T_{\rm bh}}^D.
\end{align}
The negativity of the chemical potential implies that the black hole molecules 
are attractive, which is required in order to make the black hole the stable.

\section*{Acknowledgement}

This work is supported by the National Natural Science Foundation of China under the grant
No. 12275138 and 11575088.

\section*{Data Availability Statement} 

No data was used for the research described in the article.

\section*{Declaration of competing interest}

The authors declare no competing interest.

\providecommand{\href}[2]{#2}\begingroup
\footnotesize\itemsep=0pt
\providecommand{\eprint}[2][]{\href{http://arxiv.org/abs/#2}{arXiv:#2}}



\begin{thebibliography}{99}

\bibitem{bekenstein1972black}
Bekenstein, J.D. 
``Black holes and the second law,"
\href{https://doi.org/10.1007/BF02757029}
{\emph{Lett. Nuovo Cim.} 1972, 4(15):737--740}.

\bibitem{bekenstein1973black}
Bekenstein, J.D. 
``Black holes and entropy,"
\href{https://doi.org/10.1103/PhysRevD.7.2333}{{\em Phys. Rev. D} 7(8):2333--2346, 1973.}

\bibitem{bardeen1973four}
Bardeen, J.M.; Carter, B.; Hawking, S.W. 
``The four laws of black hole mechanics,"
\href{https://doi.org/10.1007/BF01645742}{{\em Comm. Math. Phys.} 31(2):161--170, 1973.}

\bibitem{hawking1975particle}
Hawking, S.W. 
``Particle creation by black holes,"
\href{https://link.springer.com/article/10.1007/BF02345020}{
{\em Comm. Math. Phys.} 43:199-220, 1975}.

\bibitem{bekenstein1975statistical}
Bekenstein, J,D. 
``Statistical black-hole thermodynamics,"
\href{https://doi.org/10.1103/PhysRevD.12.3077}{{\em Physical Review D}, 12(10):3077, 1975}.

\bibitem{hooft1993dimensional}
't Hooft, G. 
``Dimensional reduction in quantum gravity,"
[\eprint{gr-qc/9310026}]

\bibitem{susskind1995world}
Susskind, L. 
``The world as a hologram,"
\href{https://doi.org/10.1063/1.531249}{{\em J. Math. Phys.} 36(11):6377--6396, 1995}.
[\eprint{hep-th/9409089}].

\bibitem{maldacena1998large}
Maldacena, J. 
``The large $N$ limit of superconformal field theories and
  supergravity,"
\href{https://doi.org/10.1023/A:1026654312961}{{\em Adv. Theore. \& Math. Phys.}, 
2(2):231--252, 1998}, [\eprint{hep-th/9711200}].

\bibitem{kastor2009enthalpy}
Kastor, D.; Ray, S.; Traschen, J. 
``Enthalpy and the mechanics of AdS black holes,"
\href{https://doi.org/10.1088/0264-9381/26/19/195011}{{\em Class. Quant. Grav.} 
26(19):195011, 2009.} [\eprint{0904.2765}].
 
\bibitem{dolan2010cosmological}
Dolan, B.P. 
``The cosmological constant and the black hole equation of state,"
\href{https://doi.org/10.1088/0264-9381/28/12/125020}{{\em Class. Quant. Grav.}  
28(12):125020, 2010.} [\eprint{1008.5023}].

\bibitem{dolan2011pressure}
Dolan, B.P. 
``Pressure and volume in the first law of black hole thermodynamics,"
\href{https://doi.org/10.1088/0264-9381/28/23/235017}
{{\em Class. Quant. Grav.} 28(23):235017, 2011.} [\eprint{1106.6260}].

\bibitem{dolan2011compressibility}
Dolan, B.P. 
``Compressibility of rotating black holes,"
\href{https://doi.org/10.1103/PhysRevD.84.127503}{{\em Phys. Rev. D} 84(12):127503, 2011.}
[\eprint{1109.0198}].

\bibitem{kubizvnak2012p}
Kubiz{\v{n}}{\'a}k, D.; Mann, R.B. 
``P-V criticality of charged AdS black holes,"
\href{https://doi.org/10.1007/JHEP07(2012)033}{{\em J. High Energy Phys.} 7:1--25, 2012.}
[\eprint{1205.0559}].

\bibitem{cai2013pv}
Cai, R.G.; Cao, L.M.; Li, L.; Yang, R.Q.
``P-V criticality in the extended phase space of Gauss-Bonnet black
  holes in AdS space,"
\href{https://doi.org/10.1007/JHEP09(2013)005}{{\em J. High Energy Phys.} 9:1--22, 2013.}
[\eprint{1306.6233}].

\bibitem{kubizvnak2017black}
Kubiz{\v{n}}{\'a}k, D.;  Mann, R.B.; Teo, M.
``Black hole chemistry: thermodynamics with Lambda,"
\href{https://doi.org/10.1088/1361-6382/aa5c69}{{\em Class. Quant. Grav.} 
34(6):063001, 2017.} [\eprint{1608.06147}].

\bibitem{xu2014critical}
Xu, W.; Zhao, L.
``Critical phenomena of static charged AdS black holes in conformal gravity,"
\href{https://doi.org/10.1016/j.physletb.2014.07.019}{{\em Phys. Lett. B}  
736:214--220, 2014.} [\eprint{1405.7665}].

\bibitem{xu2014gauss}
Xu, W.; Xu, H.; Zhao, L.
``Gauss--Bonnet coupling constant as a free thermodynamical variable
  and the associated criticality,"
\href{https://doi.org/10.1140/epjc/s10052-014-2970-8}
{{\em Euro. Phys. J. C} 74(7):1--13, 2014.} [\eprint{1311.3053}].


\bibitem{lemos2018black}
Lemos, J.P.S.; Zaslavskii, O.B.
``Black hole thermodynamics with the cosmological constant as
  independent variable: Bridge between the enthalpy and the Euclidean path
  integral approaches,"
\href{https://doi.org/10.1016/j.physletb.2018.08.075}
{{\em Phys. Lett. B} 786:296--299, 2018.} [\eprint{1806.07910}].

\bibitem{zhang2015phase}
Zhang, J.L.;  Cai, R.G.; Yu, H.
``Phase transition and thermodynamical geometry of
Reissner-Nordstr{\"o}m-AdS black holes in extended phase space,"
\href{https://doi.org/10.1103/PhysRevD.91.044028}
{{\em Phys. Rev. D} 91(4):044028, 2015.} [\eprint{1502.01428}].

\bibitem{wei2015insight}
Wei, S.W.; Liu, Y.X. 
``Insight into the microscopic structure of an AdS black hole from a 
thermodynamical phase transition,"
\href{https://doi.org/10.1103/PhysRevLett.115.111302}{{\em Phys. Rev. Lett.}, 
115(11):111302, 2015.} [\eprint{1502.00386}].

\bibitem{wei2020extended}
Wei, S.W.; Liu, Y.X. 
``Extended thermodynamics and microstructures of four-dimensional 
charged gauss-bonnet black hole in AdS space,"
\href{https://doi.org/10.1103/PhysRevD.101.104018}{{\em Phys. Rev. D}, 
101(10):104018, 2020.} [\eprint{2003.14275}].

\bibitem{dehyadegari2020microstructure}
Dehyadegari, A.; Sheykhi, A.; Wei, S.W. 
``Microstructure of charged AdS black hole via $P-V$ criticality,"
\href{https://doi.org/10.1103/PhysRevD.102.104013}{{\em Phys. Rev. D}, 
102(10):104013, 2020.} [\eprint{2006.12265}].

\bibitem{gao2021restricted}
Gao, Z.; Zhao, L. 
``Restricted phase space thermodynamics for AdS black holes via holography,"
\href{https://doi.org/10.1088/1361-6382/ac566c}
{{\em Class. Quant. Grav.} 39:075019, 2021.} [\eprint{2112.02386}].

\bibitem{gao2022thermodynamics}
Gao, Z.; Kong, X.; Zhao, L. 
``Thermodynamics of Kerr-AdS black holes in the restricted phase space," 
  \href{https://doi.org/10.1140/epjc/s10052-022-10080-y}
  {{\em Euro. Phys. J. C} 82(2):1--10, 2022.} [\eprint{2112.08672}].

\bibitem{wang2021black}
Wang, T.; Zhao, L. 
``Black hole thermodynamics is extensive with variable newton constant,"
\href{https://doi.org/10.1016/j.physletb.2022.136935}
{{\em Phys. Lett. B} 827:136935, 2022.} [\eprint{2112.11236}].

\bibitem{zhao2022thermodynamics}
Zhao, L. 
``Thermodynamics for higher dimensional rotating black holes with
variable Newton constant,"
\href{https://doi.org/10.1088/1674-1137/ac4f4c}
{{\em Chin. Phys. C} 46(5):055105, 2022.} [\eprint{2201.00521}].

\bibitem{kong2022restricted}
Kong, X.; Wang, T.; Gao, Z.; Zhao, L.
``Restricted phase space thermodynamics for black holes in higher
dimensions and higher curvature gravities,"
\href{https://doi.org/10.3390/e24081131}{{\em Entropy}, 24(8):1131, 2022.} 
[\eprint{2208.07748}].

\bibitem{visser2022holographic}
Visser, M.R. 
``Holographic thermodynamics requires a chemical potential for color.
\href{https://doi.org/10.1103/PhysRevD.105.106014}
{{\em Phys. Rev. D} 105(10):106014, 2022.} [\eprint{2101.04145}].

\bibitem{Gregory} Gregory, R.;  Kastor, D.;  Traschen, J.
``Black Hole Thermodynamics with Dynamical Lambda", 
\href{https://doi.org/10.1007/JHEP10%282017%29118}
{{\em JHEP} 10, 118, 2017.}  [\eprint{1707.06586}]

\bibitem{Sadeghi} Sadeghi, J.; Shokri, M.; Gashti, S. N.; Alipour, M. R. 
``RPS thermodynamics of Taub-NUT AdS black holes in the presence of 
central charge and the weak gravity conjecture,'' [\eprint{2205.03648}].

\bibitem{bwu} Bai, Y.-Y.;  Chen, X.-R.;  Xu, Z.-M.; Wu, B.
``Revisit on thermodynamics of BTZ black hole with variable Newton constant,''
[\eprint{2208.11859}].

\bibitem{G} Gibbons, G. W.;  Hawking, S. W.
``Action integrals and partition functions in quantum gravity,''
\href{https://doi.org/10.1103/PhysRevD.15.2752}{\emph{Phys. Rev. D} 15 (1977) 2752}.

\bibitem{York} York, Jr. J. W. , ``Black-hole thermodynamics and the Euclidean 
Einstein action,'' 
\href{https://doi.org/10.1103/PhysRevD.33.2092}{\emph{Phys. Rev. D} 33 (1987) 2092}.

\bibitem{Chamblin:1999tk}
Chamblin, A.; Emparan, R.; Johnson, C.~V. and Myers, R.~C. 
``Charged AdS black holes and catastrophic holography,''
\href{https://doi.org/10.1103/PhysRevD.60.064018}{{\em 
Phys. Rev. D} 60, 064018, 1999. }
[\eprint{hep-th/9902170}]. 
 
\bibitem{Gibbons2}
Gibbons, G. W. ;  Perry, M. J.;  Pope, C. N. 
``The first law of thermodynamics for Kerr-Anti-de Sitter black holes,'' 
\href{https://iopscience.iop.org/article/10.1088/0264-9381/22/9/002}
{\emph{Class. Quantum Grav.} 22 (2005) 1503}, [\eprint{hep-th/0408217}].

\bibitem{Cong1} Cong, W.; Kubiznak, D.; Mann, R.~B. 
``Thermodynamics of AdS Black Holes: Critical Behavior of the Central Charge",
\href{https://doi.org/10.48550/arXiv.2105.02223}
{{\em Phys. Rev. Lett.} 127, 9, 2021}. [\eprint{2105.02223}].

\bibitem{zhaotherm} Zhao, L. ``Statistical thermal physics,''
 \href{https://www.ecsponline.com/goods.php?id=198324}{{\em Science Press}, Beijing, 2019}, 
ISBN: 978-7-03-060735-5 (textbook in Chinese). 

\bibitem{hartnoll2009lectures}
Hartnoll, S.A.
``Lectures on holographic methods for condensed matter physics,"
\href{https://doi.org/10.1088/0264-9381/26/22/224002}{{\em Class. Quant. Grav.}, 
26(22):224002, 2009.} [\eprint{0903.3246}].

\bibitem{sachdev2011condensed}
Sachdev, S.
``Condensed matter and AdS/CFT,"
\href{https://link.springer.com/book/10.1007/978-3-642-04864-7}{In {\em From 
gravity to thermal gauge theories: the AdS/CFT 
correspondence}, pages 273--311. Springer, 2011}.

\bibitem{zaanen2015holographic}
Zaanen, J.; Liu, Y.; Sun, Y.W.; Schalm, K.
{\em Holographic duality in condensed matter physics}. 
\href{https://www.cambridge.org/core/books/holographic-duality-in-condensed-matter-physics/ED1E01756C18AE0352269202C31FE2B3}{Cambridge University Press, 2015}.

\bibitem{karch2006linear}
Karch, A.; Katz, E.; Son, D.T.; Stephanov, M.A. 
``Linear confinement and AdS/QCD,"
\href{https://doi.org/10.1103/PhysRevD.74.015005}{{\em Phys. Rev. D}, 
74(1):015005, 2006.} [\eprint{hep-ph/0602229}].

\bibitem{kajantie2007thermodynamics}
Kajantie, K.; Tahkokallio, T.; Yee, J.T. 
``Thermodynamics of AdS/QCD,"
\href{https://doi.org/10.1088/1126-6708/2007/01/019}{{\em J. High Energy Phys.}, 
2007(01):019, 2007.} [\eprint{hep-ph/0609254}].

\bibitem{panero2009thermodynamics}
Panero, M. 
``Thermodynamics of the QCD plasma and the large-N limit,"
\href{https://doi.org/10.1103/PhysRevLett.103.232001}{{\em Phys. Rev. Lett.}, 
103(23):232001, 2009.} [\eprint{0907.3719}].

\bibitem{ling2014metal}
Ling, Y.; Niu, C.; Wu, J.P.; Xian, Z.Y.; Zhang H.B.
``Metal-insulator transition by holographic charge density waves.
\href{https://doi.org/10.1103/PhysRevLett.113.091602}{{\em Phys. Rev. Lett.} 
113(9):091602, 2014}. [\eprint{1404.0777}].

\bibitem{cai2017intertwined}
Cai, R.G.; Li, L.; Wang, Y.A.; Zaanen, J.
``Intertwined order and holography: the case of parity breaking pair
  density waves,"
\href{https://doi.org/10.1103/PhysRevLett.119.181601}{{\em Phys. Rev. Lett.} 
119(18):181601, 2017}. [\eprint{1706.01470}].

\bibitem{alberte2018black}
Alberte, L.; Ammon, M.; Baggioli, M.; Jim{\'e}nez, A.;
  Pujol{\`a}s, O.
``Black hole elasticity and gapped transverse phonons in holography,"
\href{https://doi.org/10.1007/JHEP01(2018)129}{{\em J. High Energy Phys.} 
2018(1):1--30, 2018}. [\eprint{1708.08477}].

\bibitem{alberte2018holographic}
Alberte, L.; Ammon, M.; Jim{\'e}nez-Alba, A.; Baggioli, M.; Pujol{\`a}s, O.
``Holographic phonons,"
\href{https://doi.org/10.1103/PhysRevLett.120.171602}{{\em Phys. Rev. Lett.} 
120(17):171602, 2018}.  [\eprint{1711.03100}].

\end{thebibliography}

\end{document}